\documentclass[journal]{IEEEtran}
\usepackage{cite}

\ifCLASSINFOpdf
   \usepackage[pdftex]{graphicx}
\else
   \usepackage[dvips]{graphicx}
\fi
\usepackage{dcolumn} 
\usepackage{amssymb} 

\usepackage[normalem]{ulem} 

\usepackage[cmex10]{amsmath}
\usepackage{url}

\begin{document}
%
\title{Numerical methods for calculating \\poles of the scattering matrix \\with applications in grating theory}
%
%
%

\author{Dmitry A. Bykov,
        and~Leonid L. Doskolovich%
\thanks{Manuscript received July 11, 2012; revised October 12, 2012.
The work was supported by RFBR (grants \#\# 12-07-31116, 12-07-00495, 11-07-12036, 11-07-00153, 12-07-13113, 13-07-00464), by Russian Federation (RF) state contract for agreement 8027, and by RF Presidential grant (NSh-4128.2012.9) and fellowship (SP-1665.2012.5).
Copyright (c) 2012 IEEE. Personal use of this material is permitted. However, permission to use this material for any other purposes must be obtained from the IEEE by sending a request to pubs-permissions@ieee.org.%
}
\thanks{The authors are with
the Image Processing Systems Institute of the RAS, 151 Molodogvardeiskaya st., Samara, 443001, Russia
and with the Samara State Aerospace University, 34 Moskovskoye shosse, Samara, 443086, Russia
(e-mail: bykovd@gmail.com).}
}

%
%

\markboth{Journal of Lightwave Technology,~Vol.~*, No.~*, Month~2012}%
{Bykov and Doskolovich: Numerical methods for calculating poles of the scattering matrix}
%



\maketitle

\begin{abstract}
Waveguide and resonant properties of diffractive structures are often explained through the complex poles of their scattering matrices. Numerical methods for calculating poles of the scattering matrix with applications in grating theory are discussed and analyzed. A new iterative method for computing the scattering matrix poles is proposed. The method takes account of the scattering matrix form in the pole vicinity and relies upon solving matrix equations with use of matrix decompositions. Using the same mathematical approach, we also describe a Cauchy-integral-based method that allows all the poles in a specified domain to be calculated. Calculation of the modes of a metal-dielectric diffraction grating shows that the iterative method proposed has the high rate of convergence and is numerically stable for large-dimension scattering matrices. An important advantage of the proposed method is that it usually converges to the nearest pole.
\end{abstract}


%
\IEEEpeerreviewmaketitle

\section{Introduction}
%
%
%
%
\IEEEPARstart{D}{iffractive} micro- and nanostructures with resonant properties are of great interest when designing modern elements of integrated guided-wave optics and photonics (photonic crystal fibers, waveguide grating couplers, optical sensors, guided-mode resonant filters, laser resonators)~\cite{r27, r28, r29, r19, r25, r1, r9, r11,  r22,r4,r7,r23,r24,r30}. 
Appearing as an abrupt change in the transmittance and reflectance spectra, the resonant and waveguide properties are normally associated with the excitation of the structure's eigenmodes. The structure's eigenmodes can be described in terms of the complex poles of the scattering matrix~\cite{r19,r25}. Such an approach to explaining optical properties of the diffractive structures has been widely employed when describing the optical properties of diffraction gratings (2D and 3D)~\cite{r19,r25}, including those comprising anisotropic~\cite{r21} and gyrotropic~\cite{r22} materials, photonic crystal structures~\cite{r23,r24} and laser resonators~\cite{r30}.

The important practical problem requiring calculation of the poles of the scattering matrix is the design of grating/photonic-crystal waveguides with prescribed optical properties (with specified dispersion of waveguide's quasiguided or leaky modes). This design problem belongs to the class of inverse problems and is solved with use of specialized optimization techniques. When solving this inverse problem, the computationally effective methods for solving the forward problem (i.e. calculation of the scattering matrix poles) are of great practical importance.

The calculation of the scattering matrix poles is computationally challenging. A number of methods for solving this problem have been proposed in recent papers. The simplest techniques calculate the poles of the scattering matrix determinant~\cite{r23, r24} or poles of its maximal eigenvalue~\cite{r2}. A more advanced method proposed in Refs.~\cite{r9,r11,r12} relies upon the linearization of the scattering matrix inverse. In order for the optical properties of metal-dielectric diffraction gratings and plasmonic structures to be adequately described, the scattering matrix dimension should be rather high. Usually, the scattering matrix is calculated using Fourier modal method~\cite{r16}. Despite a number of approaches developed to enhance the method convergence~\cite{r17,r12}, the scattering matrix dimension can still amount to several hundreds~\cite{r32}. For scattering matrices of such large sizes operations of calculating the determinant and the scattering matrix inverse in Refs.~\cite{r24,r9,r11,r12} become numerically unstable~\cite{r13}, thus essentially limiting the methods' applicability area.

Of special note is the method that calculates the eigenmode frequencies using the Cauchy integral. This method is used when calculating modes propagating in a slab waveguides~\cite{r28,r29} and photonic crystal structures~\cite{r1,r2,r3}. It is noteworthy that this method allows all the scattering poles located in the domain of interest to be found~\cite{r1,r2,r3}.
Nevertheless, these articles made some assumptions on the form of the scattering matrix, which usually do not take place for the scattering matrix of a diffraction grating.

In this article we propose new numerical methods for calculating poles of the scattering matrix with better performance (convergence rate, computational complexity, attraction basins shape).
The paper is organized in six sections. Following the introduction, Section~{\ref{section2}} defines the scattering matrix of the diffraction grating. 
In Section~{\ref{section3}} we rigorously derive the resonant representation of the scattering matrix with use of analytic matrix-valued functions theory. The reader who is mainly interested in practical implementation of the proposed methods can skip this section. 
In Sections~{\ref{section4}},~{\ref{section4a}} we review the known methods for calculating poles of the scattering matrix. 
In Section~{\ref{section4b}} we present new iterative methods for calculating poles~[{Eqs.~(\ref{Eq19}),~(\ref{Eq20})}], which take account of the scattering matrix form in the resonance vicinity. 
In Section~{\ref{section4c}} a new formulation of the Cauchy-integral-based method~[Eq.~{(\ref{Eq26})}]
 is proposed. In Section~{\ref{section5}} we conduct the numerical comparison of the existing methods with the proposed ones.

\section{\label{section2}Scattering matrix}
\begin{figure}[b]
\centering
\includegraphics{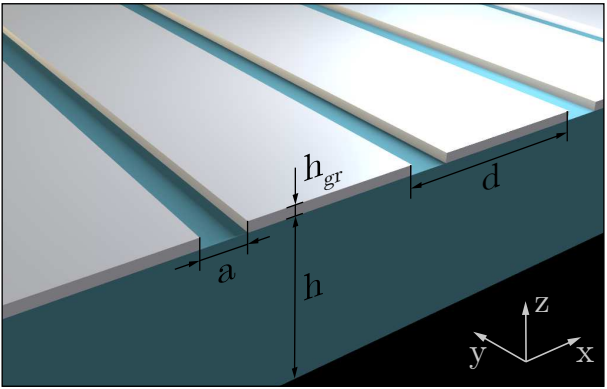}
\caption{\label{fig:fig1}(Color online) Silver diffraction grating on dielectric waveguide-layer (grating parameters: period $d=1000\; {\rm nm}$, slit depth $h_{\rm gr} =50\; {\rm nm}$, slit width $a=200\; {\rm nm}$, waveguide layer thickness $h=800\; {\rm nm}$, layer's permittivity $\varepsilon = 5.5$).}
\end{figure}

Let us give the definition of the scattering matrix for the case of 2D diffraction grating. Assume that the grating is periodic along the $x$-axis, with period $d$ (see Fig.~\ref{fig:fig1}). Let there be a set of plane waves incident on the grating from superstrate and substrate, with the $x$-component of the incident light wave-vectors given by 
\begin{equation}
\label{Eq1}
k_{x,m} = k_{x,0} +\frac{2\pi }{d} m,\, \, \, m\in \mathbb{Z}.
\end{equation} 

Due to diffraction by the grating, the set of incident plane waves is scattered into a set of reflected and transmitted diffraction orders. Note that according to Floquet theorem, the $x$-components of the diffraction orders' wave-vectors are also defined by Eq.~(\ref{Eq1}). 

 The grating scattering matrix $S$ relates the complex amplitudes of the incident ($\Psi^{\rm inc}$) and scattered ($\Psi^{\rm scatt}$) waves by the formula: 
\begin{equation}
\label{Eq2} 
S^{-1} \Psi^{\rm scatt} =\Psi^{\rm inc},
\end{equation} 
where $\Psi ^{\rm inc} =\left[\begin{array}{c} I_1 \\ I_2 \end{array}\right]$, $\Psi ^{\rm scatt} =\left[\begin{array}{c} R \\ T \end{array}\right],$ $R$ and $T$ are the complex amplitude vectors of the reflected and transmitted diffraction orders, whereas $I_1$ and $I_2$ are the complex amplitude vectors of the waves incident on the grating from superstrate and substrate. Thus, the scattering matrix element $(S)_{ij}$ is the complex scattering amplitude of the $j$-th incident wave in the direction of the $i$-th scattered wave. Note that the diffraction of the incident waves by the grating produces an infinite number of propagating and evanescent orders, so that the matrix $S$ also has infinite dimensions. To conduct the numerical simulation, the scattering matrix is truncated. Let $N=2K+1$ denote the truncation order, which corresponds to the incident and scattered waves with the numbers $m=-K, \ldots, K$ in Eq.~(\ref{Eq1}). Then, the scattering matrix relates $2N$ incident and $2N$ scattered waves. Considering that the incident and scattered waves have two polarization states, the scattering matrix dimension is $4N\times 4N$. 
The calculation of the truncated scattering matrix can be carried out using the Fourier modal method~{\cite{r16}} also referenced as the Scattering matrix method~{\cite{r19}}.
In general, the scattering matrix can be calculated for multilayer coatings, 2D and 3D periodic structures as well as for aperiodic structures (with use of PML or absorbing layers).

For a given geometry and materials of the grating, the scattering matrix $S$ is a function of frequency $\omega$ and the $x$-component of the wave-vector of the incident wave with number $m=0$: $S=S\left(\omega ,k_{x,0} \right)$. When describing the resonances in the grating transmittance and reflectance spectra, the incident wave direction is specified and the scattering matrix is treated as a function of frequency $\omega$: $S=S\left(\omega \right)$. In this case the structure's resonances are found as the poles of the analytic continuation of matrix $S\left(\omega \right)$~\cite{r9}. The real part of a pole corresponds to the frequency of the incident wave that can excite the corresponding eigenmode, while the inverse of the imaginary part defines the lifetime of the resonance.

\section{\label{section3}Resonance representation of the scattering matrix}
In this section we rigorously derive the resonant representation of the scattering matrix~{(\ref{Eq9})}
with use of analytic matrix-valued functions theory. We believe that this section is important, because it gives a rigorous basis for the methods proposed in Section~{\ref{section4}}. Nevertheless, the reader interested mainly in the practical application of the numerical methods can skip current section and keeping Eq.~{(\ref{Eq9})}
 in mind go to Section~{\ref{section4}}.

Consider the analytic continuation $S(\omega),\, \omega \in \mathbb{C}$ of the scattering matrix onto a complex $\omega$-plane region $D$ bounded by a closed curve $\Gamma$. We assume that the Rayleigh anomalies are located far away from the frequency range of interest. Then, the analytic continuation $S(\omega )$ in the region $D$ will be single-valued~\cite{r27,r26}.

Assume that in the $D$ region there is a simple pole of the scattering-matrix analytic continuation at $\omega = \omega_{\rm p}$. Below, only the simple poles of the scattering matrix are considered. This assumption is widely used in scattering theory both in electrodynamics and quantum mechanics~{\cite{r1}}. In this case, it makes sense to define the $S(\omega)$ matrix residue: 
\begin{equation}
\label{Eq3}
\mathop{\operatorname{Res}}\limits_{\omega =\omega_{\rm p}} S(\omega )=\frac{1}{2\pi {\rm i}} \oint_{\gamma}S(\omega )\, {\rm d}\omega,
\end{equation} 
where the integration contour $\gamma$ is chosen in such a way as to contain just a single pole $\omega_{\rm p}$. Equation~(\ref{Eq3}) should be understood as an element-wise operation. If the scattering matrix has a single pole in the region $D$, the following relation takes place:
\begin{equation}
\label{Eq4}
S(\omega) = A(\omega)+\frac{B}{\omega -\omega_{\rm p} },
\end{equation} 
where $B=\operatorname{Res}\nolimits_{\omega = \omega_{\rm p}} S(\omega)$ and the matrix-valued function $A(\omega)$ has no poles in the region $D$, thus being holomorphic in this region. In the general case of $M$ poles found in the region $D$, the decomposition in Eq.~(\ref{Eq4}) takes the form:
\begin{equation}
\label{Eq5}
S(\omega) = A\left(\omega \right)+\sum _{m=1}^{M}\frac{B_m}{\omega -\omega_{\rm p}^{(m)} },
\end{equation}
where $B_m=\operatorname{Res}\nolimits_{\omega =\omega_{\rm p}^{(m)} } S(\omega)$. The first term in Eqs.~(\ref{Eq4}) and~(\ref{Eq5}) describes the non-resonant scattering of light and the second --- the resonant scattering. 

Let us now discuss the matrix properties of $S(\omega)$ in the context of scattering (diffraction) theory. Assume that $S(\omega)$ has a simple pole at $\omega = \omega_{\rm p}$, while the inverse matrix $S^{-1} (\omega)$ elements have no pole at this frequency. In this case, the kernel of the $S^{-1} (\omega_{\rm p})$ matrix defines non-trivial solutions of the homogeneous equation:
\begin{equation}
\label{Eq6}
S^{-1} \Psi ^{\rm scatt} = 0.
\end{equation} 
Thus, $\ker S^{-1} (\omega_{\rm p})$ describes the field distribution in the absence of incident waves, i.e. the frequency $\omega_{\rm p}$ corresponds to the quasiguided mode of the grating.

It can easily be shown that $\operatorname{Im}B=\ker S^{-1} (\omega_{\rm p})$ and hence, $\operatorname{rank}\, B=\dim \ker S^{-1} (\omega_{\rm p})$~\cite{r1}. As a rule, $\operatorname{rank} B=1$, i.e. only one mode corresponds to the frequency $\omega_{\rm p} $. However, at certain grating parameters, the frequencies of several different modes may coincide. In this case, $\operatorname{rank} B>1$. Such resonances are referred to as degenerate. The degenerate-resonance structures have interesting optical properties which can be applied to the design of channel drop filters and all pass filters~\cite{r4,r7}.

In the general case, assume that $\operatorname{rank}B=r$. Then, for the matrix $B\in \mathbb{C}^{n\times n} $, a rank factorization can be written~\cite{r8}:
\begin{equation}
\label{Eq7}
B=LR,
\end{equation} 
where $L\in \mathbb{C}^{n\times r} $, $R\in \mathbb{C}^{r\times n} $, $\operatorname{rank}L=\operatorname{rank}R=r$. In view of Eq.~(\ref{Eq7}), Eq.~(\ref{Eq4}) can be represented as
\begin{equation}
\label{Eq8}
S=A(\omega )+L\frac{1}{\omega -\omega_{\rm p}} R.
\end{equation}
Accordingly, the general decomposition in Eq.~(\ref{Eq5}) takes the form:
\begin{equation}
\label{Eq9}
S=A(\omega )+\sum_{m=1}^{M}L_m \frac{1}{\omega -\omega_{\rm p}^{(m)}} R_m  =A(\omega)+L(I\omega -\Omega _{\rm p})^{-1} R,
\end{equation}
where $L$ and $R$ are the block matrices defined as $L=\left[\begin{array}{cccc} L_1 & L_2 & \cdots & L_M \end{array}\right]$, $R^{\rm T} =\left[\begin{array}{cccc} R_{1}^{\rm T} & R_2^{\rm T} & \cdots & R_M^{\rm T} \end{array}\right]$, and $\Omega_{\rm p}$ is the diagonal matrix composed of $\omega _{{\rm p}}^{(m)} ,\, m=1,...,M$, with the frequency $\omega _{{\rm p}}^{(m)}$ recurring as many times as is the rank of the corresponding matrix $B_{m} $. Equations~(\ref{Eq8}),~(\ref{Eq9}) define resonance representations of the scattering matrix which will be used in further developments of numerical methods for computing the scattering matrix poles.

\section{\label{section4}Computing the scattering matrix poles}
As demonstrated above, the poles of the scattering matrix define the grating eigenmodes. Consider a problem of computing the scattering matrix $S(\omega )$ poles located in the region $D$. The simplest approach assumes that the poles of the matrix-valued function $S(\omega )$ can be found as the poles of its determinant, thus by solving numerically the following equation:
\begin{equation}
\label{Eq10}
1/\det S(\omega )=0.
\end{equation}
This approach allows the poles of a small-dimension matrix $S(\omega )$ to be found. For a large-dimension matrix $S(\omega )$, the calculation of the determinant becomes numerically unstable. Showing a higher stability is the problem of solving the equation~\cite{r2}:
\begin{equation}
\label{Eq11}
1/\max \operatorname{eig}S(\omega )=0,
\end{equation} 
where $\max \operatorname{eig}S(\omega )$ is the maximal-modulus eigenvalue of the matrix $S(\omega )$. Equations~(\ref{Eq10}) and~(\ref{Eq11}) can be solved with iterative methods for deriving the root of a nonlinear equation, for instance, Newton's method or, more general, Householder's method~\cite{r31}. For Eq.~(\ref{Eq11}), this method is given by the following iterative procedure:
\begin{equation}
\label{Eq12}
\omega_{n+1} =\omega_n +p\left. \frac{\frac{{\rm d}^{p-1} }{{\rm d}\omega^{p-1} } \max \operatorname{eig}S(\omega )}{\frac{{\rm d}^p }{{\rm d}\omega^p } \max \operatorname{eig}S(\omega )} \right|_{\omega = \omega_n},
\end{equation} 
where $\omega_n$ is the initial pole guess. At $p=1$, Eq.~(\ref{Eq12}) corresponds to Newton's method; at $p=2$ --- to Halley's method. When solving Eqs.~(\ref{Eq10}) and~(\ref{Eq11}), the method of Eq.~(\ref{Eq12}) disregards the form of the scattering matrix in Eq.~(\ref{Eq9}) replacing it with a scalar value [determinant or maximal-modulus eigenvalue of $S(\omega)$].
This is the reason of rather slow convergence of the above mentioned methods.
In the following subsections we consider the iterative methods for computing the scattering matrix poles with the use of matrix decompositions.

\subsection{\label{section4a}Computing the poles through linearization of the scattering matrix inverse}
Let us analyze the iterative method for computing the scattering matrix poles proposed in Refs.~\cite{r9,r11,r12,r13}. Let $\omega_n$ be an initial pole guess. We decompose the matrix $S^{-1} (\omega )$ into a Taylor series up to the first term:
\begin{equation}
\label{Eq13}
S^{-1} (\omega) = S^{-1} (\omega_n)+\left. \frac{{\rm d}S^{-1} }{{\rm d}\omega} \right|_{\omega_n} (\omega -\omega_n ). 
\end{equation} 
Assume that $\omega_{\rm p}$ is the scattering matrix pole, then there exists a vector $\Psi^{\rm scatt}$ that represents a non-trivial solution of the system~(\ref{Eq6}). Multiplying Eq.~(\ref{Eq13}) by $\Psi^{\rm scatt}$ on the right at $\omega =\omega_{\rm p}$ yields:
\begin{equation}
\label{Eq14}
S^{-1} (\omega_n )\Psi^{\rm scatt} =(\omega_n -\omega_{\rm p} )\left. \frac{{\rm d}S^{-1} }{{\rm d}\omega} \right|_{\omega_n} \Psi ^{\rm scatt}.
\end{equation}
Equation~(\ref{Eq14}) defines a generalized eigenvalue problem. Solving this problem yields a set of eigenvalues $\lambda_k =\omega_n -\omega_{\rm p}$. Choosing out of these a minimal-modulus eigenvalue, we obtain the following iterative method:
\begin{equation}
\label{Eq15}
\omega_{n+1} = \omega_n - \min \operatorname{eig}\left(S^{-1} (\omega_n ),\left. \frac{{\rm d}S^{-1} }{{\rm d}\omega } \right|_{\omega_n} \right), 
\end{equation}
where $\operatorname{eig}(F,G)$ denotes a vector composed of the eigenvalues for the generalized eigenvalue problem, $F X=\lambda \, G X$. Choosing in Eq.~(\ref{Eq15}) a minimal-modulus eigenvalue means that the subsequent guess $\omega_{n+1}$ chosen will be closest to the initial guess $\omega_n $.

Note that the method of Eq.~(\ref{Eq15}) has a number of disadvantages, the major being inability to conduct the numerically stable computation of the scattering matrix inverse for a large number of diffraction orders~\cite{r13}. Another disadvantage is that if a pole of the matrix $S^{-1}(\omega)$ is located in the vicinity of the pole of the scattering matrix $S(\omega)$, the Taylor series~(\ref{Eq13}) will diverge (see Appendix~\ref{appendixB}).

\subsection{\label{section4b}Pole calculation based on a resonance approximation}
Let us consider a new iterative technique for scattering matrix pole calculation that takes account of the scattering matrix form in the resonance vicinity. Assume that an initial pole guess is $\omega=\omega_n$. Put down the resonance approximation~(\ref{Eq9}) neglecting the frequency-dependence of the non-resonance term:
\begin{equation}
\label{Eq16}
S(\omega)=A+L(\omega I-\Omega_{\rm p} )^{-1} R.
\end{equation} 

The first and second derivatives of $S(\omega )$ are
\begin{equation}
\label{Eq17}
\begin{aligned}
S'(\omega)&=-L(\omega I-\Omega_{\rm p})^{-2} R, 
\\
S''(\omega)&=2L(\omega I-\Omega_{\rm p})^{-3} R.
\end{aligned} 
\end{equation} 

Assume that $\operatorname{rank}L=\operatorname{rank}R=\sum_m \operatorname{rank}L_m =\sum_m \operatorname{rank}R_m $. This assumption implies that the columns of the matrix $L$ are linearly independent or, which is the same, the kernels of the matrices $S^{-1} (\omega _{\rm p}^{(m)} ),\, \, m=1,\ldots ,M$ are linearly independent. The latter means that the scattered field distributions for different modes are linearly independent. This assumption normally holds when the scattering matrix dimension is much larger than the number of modes ($\dim S>\operatorname{rank}L$).

Putting $\omega = \omega_{n}$ in Eq.~(\ref{Eq17}) yields a system of two matrix equations with respect to an unknown diagonal matrix $\Omega_{\rm p}$. A method for solving equations of this kind is considered in Appendix~\ref{appendixA}. Following that method [see Eq.~(\ref{EqA6})], the diagonal matrix $\Omega_{\rm p}$ is given by
\begin{equation}
\label{Eq18}
\Omega_{{\rm p}} =\omega _{n} I{\rm +2}\operatorname{diag}\operatorname{eig}(U_r^{\dagger} S'(\omega _{n} )V_r \Sigma _r^{-1} ),
\end{equation}
where $\operatorname{diag}\operatorname{eig}\, F$ is the diagonal matrix composed of the eigenvalues of the matrix $F$ and the matrices $U_r ,\, \, \Sigma _r ,\, \, V_r$ are derived from the compact singular value decomposition of the matrix $S''(\omega_n) = U_r \Sigma_r V_r^{\dagger}$. From Eq.~({\ref{Eq18}}), we can obtain the following iterative method:
\begin{equation}
\label{Eq19}
\omega_{n+1} = \omega_n + 2\, \min \operatorname{eig}\left(U_r^{\dagger} S'(\omega_n) V_r \Sigma_r^{-1} \right).
\end{equation}
In a similar way to Eq.~(\ref{Eq15}), taking the eigenvalue with the minimal modulus in Eq.~(\ref{Eq19}) denotes that the subsequent pole guess chosen $\omega _{n+1}$ will be the pole closest to the initial guess $\omega_n$.

The iterative method in Eq.~(\ref{Eq19}) assumes that there are several poles in the initial guess vicinity. In practice, one can assume that in the vicinity of $\omega _{n}$ there is a single pole corresponding to the non-degenerate resonance. In this case, in view of Eq.~(\ref{EqA9}), the iterative method takes a simple form:
\begin{equation}
\label{Eq20}
\omega_{n+1} =\omega _{n} +2\frac{\max \operatorname{eig}S'(\omega_n)}{\max \operatorname{eig}S''(\omega_n)}.
\end{equation} 

As distinct from~(\ref{Eq15}), the iterative methods in Eqs.~(\ref{Eq19}) and~(\ref{Eq20}) are based on the resonance approximation~(\ref{Eq16}), rather than the linearization of the matrix $S^{-1} (\omega )$. The results of the numerical investigation in Section~\ref{section5} demonstrate that the iterative method~(\ref{Eq19}) and~(\ref{Eq20}) show a better convergence when compared with Eq.~(\ref{Eq15}). Besides, the benefit of the proposed approach is that it remains valid for a large number of diffraction orders (large-dimension matrix S) and when the poles of the scattering matrix and of its inverse are close to each other.

Note that the method of Eq.~(\ref{Eq15}) has the meaning of Newton's method for the matrix-valued functions, whereas the methods of Eqs.~(\ref{Eq19}) and~(\ref{Eq20}) can be treated as a matrix extension of Halley's method for solving equations of the form $1/f(x)=0$. In general, it is possible to write down a matrix analog of Householder's method~\cite{r31} with the aid of the $p$-th and ($p-1$)-th derivatives of the scattering matrix: $S^{(p)}(\omega_n)$, $S^{(p-1)}(\omega_n)$. In this case, the iterative method of Eq.~(\ref{Eq19}) will represent a particular case (at $p=2$) of the following iterative method:
\begin{equation}
\label{Eq21}
\omega_{n+1} =\omega_{n} +p\min \operatorname{eig}\left(U_r^{\dagger} S^{(p-1)} \left(\omega _{n} \right)V_r \Sigma_r^{-1} \right),
\end{equation}
where the matrices $U_r ,\Sigma _r ,V_r$ are derived from the compact singular value decomposition of the matrix $S^{(p)}(\omega_n) = U_r \Sigma _r V_r^{\dagger} $. The analog of the method in Eq.~(\ref{Eq20}) is written in a similar way:
\begin{equation}
\label{Eq22}
\omega_{n+1} =\omega _{n} +p\frac{\max \operatorname{eig}S^{(p-1)} (\omega _{n} )}{\max \operatorname{eig}S^{(p)} (\omega _{n} )}.
\end{equation} 
The above relation resembles Householder's method of Eq.~(\ref{Eq12}) for solving Eq.~(\ref{Eq11}), except that the operations of differentiation and calculation of the maximal-modulus eigenvalue are swapped. At $p=1$, the methods in Eq.~(\ref{Eq21}) and~(\ref{Eq22}) represent analogs of Newton's method, but, as distinct from Eq.~(\ref{Eq15}), the said method relies upon the calculation of the scattering matrix $S(\omega )$ and its derivative, rather than of matrix $S^{-1}(\omega)$.

\subsection{\label{section4c}Pole calculations based on the Cauchy integral }
The efficiency of the above-discussed methods in Eqs.~(\ref{Eq10}),~(\ref{Eq11}),~(\ref{Eq15}),~(\ref{Eq19}), and~(\ref{Eq20}) essentially depends on the initial pole guess $\omega =\omega _{n} $. Besides, the said methods allow one to find just a single pole in the initial guess' vicinity. Note that with methods of Eqs.~(\ref{Eq15}),~(\ref{Eq19}) and~(\ref{Eq21}) it is possible to use several eigenvalues when calculating the subsequent approximations. However, such an approach does not guarantee that \emph{all} the poles of the scattering matrix will be found in the $D$ region of interest.

Below, we discuss an approach based on the Cauchy integral that enables all the scattering matrix poles to be found in the region of interest. Such an approach is widely utilized when calculating modes of multilayered slab waveguides~\cite{r28,r29} and photonic-crystal waveguides~\cite{r1,r2}.

Following Refs.~\cite{r1,r2,r3}, calculate two contour integrals with respect to the matrix-valued functions:
\begin{equation}
\label{Eq23}
\begin{aligned}
C_1 &= \frac{1}{2\pi {\rm i}} \oint_{\Gamma}S(\omega)\;{\rm d}\omega;
\\
C_2 &=\frac{1}{2\pi {\rm i}} \oint_{\Gamma}\omega S(\omega)\;{\rm d}\omega, 
\end{aligned}
\end{equation} 
where $\Gamma$ is the boundary of the region $D$. Making use of Eq.~(\ref{Eq5}), we easily find that
\begin{equation}
\label{Eq24}
\begin{aligned}
C_1 &= \sum_m B_m;
\\
C_2 &= \sum_m\omega_{\rm p}^{(m)} B_m,
\end{aligned}
\end{equation} 
where $B_m = \operatorname{Res}_{\omega = \omega_{\rm p}^{(m)}} S(\omega)$. Note that in Ref.~\cite{r1} it was demonstrated with a number of examples that all the poles found within the contour $\Gamma$ could be derived using the integrals~(\ref{Eq23}). It should be noted that in Ref.~\cite{r1} the matrices $B_m$ are assumed to be symmetrical and the columns of the matrix $L$ in Eq.~(\ref{Eq9}) --- orthogonal. The said assumptions are usually not valid for the diffraction grating scattering matrices~\cite{r11}.

Let us discuss the method for solving Eq.~(\ref{Eq24}) under more general assumptions. To these ends, Eq.~(\ref{Eq24}) is written using the notations of Eq.~(\ref{Eq9}):
\begin{equation}
\label{Eq25}
\begin{aligned}
C_1 &= LR; \\
C_2 &= L\Omega_{\rm p} R.
\end{aligned}
\end{equation} 
As earlier, the columns of the matrix $L$ are assumed to be linearly independent. This condition is more general than the orthogonality condition employed in Ref.~\cite{r1}. The system of equations~(\ref{Eq25}) will be solved with respect to the unknown diagonal matrix $\Omega_{\rm p}$ using a method discussed in Appendix~\ref{appendixA}. Following this method the $\Omega_{\rm p}$ matrix is given by
\begin{equation}
\label{Eq26}
\Omega _{\rm p} = \operatorname{diag}\operatorname{eig}(U_r^{\dagger} C_2 V_r \Sigma_r^{-1} ), 
\end{equation} 
where $U_r ,\, \, \Sigma _r ,\, \, V_r$ are derived from the compact singular value decomposition of the matrix $C_1 = U_r \Sigma_r V_r^{\dagger}$.

Thus, calculating the integrals in Eq.~(\ref{Eq23}) enables all the scattering matrix poles located inside the integration contour to be found. The pole calculation accuracy depends on the accuracy of the numerical calculation of the integrals~(\ref{Eq23}). To calculate the poles with high accuracy, a large number of the scattering matrix calculations needs to be conducted~\cite{r1,r2}. Because of this, it is relevant to use the Cauchy-integral-based method only when calculating the initial pole guess, with the subsequent highly accurate values to be derived using the above-considered iterative techniques.

\section{\label{section5}Numerical examples}
In this section we study the performances of the above-discussed methods for computing the poles of the diffraction grating scattering matrix. The diffraction grating scattering matrix is calculated using the Fourier modal method~\cite{r16,r17,r18}. Note that the condition of analyticity of the scattering matrix requires proper implementation of the Fourier modal method in the case of complex frequencies~\cite{r19,r12}.

Let us consider the calculation of the eigenmodes of the diffraction grating composed of a periodic array of slits in a silver film coated on a dielectric waveguide layer (Fig.~\ref{fig:fig1}). We are interested in the modes that can be excited by the normally incident plane wave ($k_{x,0} = 0$). The grating parameters are given in the caption to Fig.~\ref{fig:fig1}. The permittivity of silver is described using the Lorentz--Drude model~\cite{r20}, with the analytical approximation for $\varepsilon_{\rm Ag}(\omega)$ being treated as a function of complex frequency. Note that the analyticity of the function $\varepsilon(\omega)$ is required to provide the analyticity of the scattering matrix $S(\omega )$. If this condition is violated, the convergence rate of the iterative techniques will be essentially deteriorated, while the Cauchy-integral-based method will become inapplicable. Below, the grating mode frequency $\omega_{\rm p}$ is assumed to have been calculated correctly if $\max \operatorname{svd} S(\omega_{\rm p}) \ge 10^{10} $, where $\max\operatorname{svd}$ denotes maximal singular value. This condition is also used as a termination criterion for iterative methods. The numerical calculations show that for the grating under study the above condition corresponds to the pole determination accuracy of $|\Delta \lambda_{\rm p} | < 10^{-8} \, {\rm nm}$. Such an accuracy is sufficient for practical uses, including the calculation of the mode field distribution.

\subsection{Cauchy-integral-based method}
Let us determine the entire set of modes supported by the grating in the frequency range $\omega \in [1.0\times 10^{15} \;{\rm s}^{-1} ;1.8\times 10^{15} \;{\rm s}^{-1} ]$. Note that only high quality modes are considered ($|\operatorname{Im}\omega_{\rm p} |<5\times 10^{12} \; {\rm s}^{-1} $). We used the value of $N=21$ as the truncation order for subsequent calculations.

For estimating the mode frequencies, we calculated the contour integrals of Eq.~(\ref{Eq23}) using the trapezoidal method with 500 discretization points. Ten approximate pole values were derived from Eq.~(\ref{Eq26}). The said values are shown in column~2 of Table~\ref{tab:table1}. The accurate values of the pole frequencies calculated by the iterative method of Eq.~(\ref{Eq20}) are shown in column~3 of Table~\ref{tab:table1}. The initial guess error is presented in column~4. Table~\ref{tab:table1} suggests that with the contour integration method, the poles can be calculated with a sufficiently high accuracy.

\begin{table*}
\caption{\label{tab:table1}Poles derived from Eq.~(\ref{Eq26}) and a more accurate iterative pole calculation.}
\centering
\begin{tabular}{cD{.}{.}{27}D{.}{.}{27}D{.}{.}{1.9}ccc} 
\hline
&  &  &  
& \multicolumn{3}{c}{$n_{\rm iter} $} \\ \cline{5-7} 
& \multicolumn{1}{c}{$\omega_{\rm Cauchy}$} 
& \multicolumn{1}{c}{$\omega_{\rm accurate}$} 
& \multicolumn{1}{c}{$\left|\omega_{\rm accurate} -\omega_{\rm Cauchy} \right|$} 
& \multicolumn{1}{c}{\parbox[t]{1.3cm}{$ $\\Eq.~(\ref{Eq20})}} 
& \multicolumn{1}{c}{\parbox[t]{1.3cm}{$ $\\Eq.~(\ref{Eq15})}} 
& \multicolumn{1}{c}{\parbox[t]{1.3cm}{Newton\\for Eq.~(\ref{Eq10})}} \\ \hline
$\omega_{1}$  & 1.14246\times 10^{15} -3.35055\times 10^{12} {\rm i} & 1.14247\times 10^{15} -3.34402\times 10^{12} {\rm i} & 7.86650\times 10^{9} & 1 & 2 & 3 \\ 
$\omega_{2}$  & 1.15554\times 10^{15} -6.85407\times 10^{11} {\rm i} & 1.15551\times 10^{15} -6.86647\times 10^{11} {\rm i} & 3.40401\times 10^{10} & 2 & 2 & 3 \\
$\omega_{3}$  & 1.40431\times 10^{15} -1.87394\times 10^{12} {\rm i} & 1.40428\times 10^{15} -1.87643\times 10^{12} {\rm i} & 3.45275\times 10^{10} & 2 & 2 & 3 \\
$\omega_{4}$  & 1.48168\times 10^{15} -3.86736\times 10^{12} {\rm i} & 1.48168\times 10^{15} -3.86743\times 10^{12} {\rm i} & 2.35895\times 10^{9} & 1 & 2 & 3 \\
$\omega_{5}$  & 1.50470\times 10^{15} -1.11064\times 10^{12} {\rm i} & 1.50469\times 10^{15} -1.11116\times 10^{12} {\rm i} & 1.52001\times 10^{10} & 2 & 2 & 3 \\
$\omega_{6}$  & 1.66425\times 10^{15} -7.11868\times 10^{11} {\rm i} & 1.66425\times 10^{15} -7.12688\times 10^{11} {\rm i} & 3.44210\times 10^{9} & 1 & 2 & 3 \\
$\omega_{7}$  & 1.66551\times 10^{15} -1.05160\times 10^{11} {\rm i} & 1.66545\times 10^{15} -1.06986\times 10^{11} {\rm i} & 5.64199\times 10^{10} & 2 & 2 & 4 \\
$\omega_{8}$  & 1.73171\times 10^{15} -3.70973\times 10^{12} {\rm i} & 1.73171\times 10^{15} -3.70850\times 10^{12} {\rm i} & 1.23369\times 10^{9} & 1 & 2 & 3 \\
$\omega_{9}$  & 1.73768\times 10^{15} -2.43030\times 10^{12} {\rm i} & 1.73768\times 10^{15} -2.43055\times 10^{12} {\rm i} & 2.84178\times 10^{9} & 1 & 2  & 3 \\
$\omega_{10}$ & 1.77660\times 10^{15} -2.79575\times 10^{12} {\rm i} & 1.77659\times 10^{15} -2.79747\times 10^{12} {\rm i} & 6.09846\times 10^{9} & 1 & 2  & 3 \\
\hline
\end{tabular}
\end{table*}

Note, for comparison, that if the pole approximations were obtained by calculating the scattering matrix on a $500\times 20$ grid (i.e. 10000 calculations of the scattering matrix), we would be able to find only nine poles. The reason is that the poles~$\omega_6$,~$\omega_7$ are located very close to each other. Thus, the use of the Cauchy-integral-based method is appropriate in this case.

\subsection{Comparison of the convergence rates of the iterative methods}
Consider the convergence of the iterative methods of Eqs.~(\ref{Eq15}),~(\ref{Eq20}), and Newton's method when solving Eq.~(\ref{Eq10}). First, analyze the situation when fairly good initial guesses of the poles are known. In this case, the pole estimates shown in column~2 of Table~\ref{tab:table1} can be refined using the methods under study. Columns~5,~6,~7 of Table~\ref{tab:table1} give the number of iterations when calculating the poles by the methods~(\ref{Eq20}),~(\ref{Eq15}), and Newton's method when solving Eq.~(\ref{Eq10}). From Table~\ref{tab:table1} it follows that the convergence rate is lowest for Newton's method when solving Eq.~(\ref{Eq10}). Meanwhile, the proposed method of Eq.~(\ref{Eq20}) and the method of Eq.~(\ref{Eq15}) require about the same number of iterations.

Let us investigate in more detail the convergence rate when the initial pole guesses are not known. To do so, the following numerical experiment was conducted. For a random point from the complex frequency range under study, the iterative method was initiated. Tables~\ref{tab:table2},~\ref{tab:table3},~\ref{tab:table4} give the average percentage of the initial guesses having converged to the pole ($n_{\rm conv}$), the average number of iterations ($n_{\rm iter}$), and the average number of the scattering matrix calculations ($n_{\rm s-calc}$).

\begin{table}
\begin{minipage}{\columnwidth}
\caption{\label{tab:table2}Convergence of the iterative methods for the truncation order $N=21$ ($84\times 84$ scattering matrix).}
\centering
\renewcommand*{\thefootnote}{\alph{footnote}} 
\begin{tabular}{cD{.}{.}{3.1}D{.}{.}{1.2}D{.}{.}{2.2}D{.}{.}{2.1}}
\hline
Method & \multicolumn{1}{c}{$n_{\rm conv}$} ,\% & \multicolumn{1}{c}{$n_{\rm iter}$} & \multicolumn{1}{c}{$n_{\rm s-calc}$} & \multicolumn{1}{c}{$n_{\rm near}$} ,\% \\ \hline
Newton for Eq.~(\ref{Eq10})\footnotemark[1] & 98.5 & 9.85 & 19.7  & 57.2 \\
Newton for Eq.~(\ref{Eq11})\footnotemark[1] & 94.9 & 7.08 & 14.16 & 42.4 \\
Halley for Eq.~(\ref{Eq11})\footnotemark[1] & 99.9& 4.20 & 12.60 & 54.5 \\
Eq.~(\ref{Eq15})\footnotemark[2] & 99.7 & 5.48 & 10.96 & 63.9 \\
Eq.~(\ref{Eq19})\footnotemark[2] & 100  & 2.78 & 8.35  & 96.1 \\
Eq.~(\ref{Eq20})\footnotemark[2] & 100  & 3.40 & 10.2  & 64.6 \\
\hline
\end{tabular}
\footnotetext[1]{Scalar method}
\footnotetext[2]{Matrix method}
\end{minipage}

\caption{\label{tab:table3}Convergence of the iterative methods for the truncation order $N=61$ ($244\times 244$ scattering matrix).}
\centering
\begin{tabular}{cD{.}{.}{3.1}D{.}{.}{2.2}D{.}{.}{2.2}D{.}{.}{2.2}}
Method & \multicolumn{1}{c}{$n_{\rm conv}$} ,\% & \multicolumn{1}{c}{$n_{\rm iter}$} & \multicolumn{1}{c}{$n_{\rm s-calc}$} & \multicolumn{1}{c}{$n_{\rm near}$} ,\% \\ \hline
 Newton for Eq.~(\ref{Eq10}) &  82.5 & 11.79 & 23.58 & 37.7 \\
 Newton for Eq.~(\ref{Eq11}) & 87.2& 7.1 & 14.2 & 24.5 \\
 Halley for Eq.~(\ref{Eq11}) & 99.7& 4.54 & 13.62 & 40.3 \\
 Eq.~(\ref{Eq15}) & 78 & 5.95 & 11.9 & 48.6 \\
 Eq.~(\ref{Eq19}) & 100  & 2.88 & 8.64  & 87.6 \\
 Eq.~(\ref{Eq20}) & 100  & 3.72 & 11.16  & 43.75 \\
\hline
\end{tabular}

\caption{\label{tab:table4}Convergence of the iterative methods for the truncation order $N=201$ ($804\times 804$ scattering matrix).}
\centering
\begin{tabular}{cD{.}{.}{3}D{.}{.}{1.2}D{.}{.}{2.2}D{.}{.}{2.1}}
\hline
Method & \multicolumn{1}{c}{$n_{\rm conv}$} ,\% & \multicolumn{1}{c}{$n_{\rm iter}$} & \multicolumn{1}{c}{$n_{\rm s-calc}$} & \multicolumn{1}{c}{$n_{\rm near}$} ,\% \\ \hline
 Newton for Eq.~(\ref{Eq10}) &  0 & \multicolumn{1}{c}{---} & \multicolumn{1}{c}{---} & \multicolumn{1}{c}{---} \\
 Newton for Eq.~(\ref{Eq11}) & 94& 6.83 & 13.66 & 26.5 \\
 Halley for Eq.~(\ref{Eq11}) & 99& 4.85 & 14.55 & 40 \\
 Eq.~(\ref{Eq15}) & 0 & \multicolumn{1}{c}{---} & \multicolumn{1}{c}{---} & \multicolumn{1}{c}{---} \\
 Eq.~(\ref{Eq19}) & 100  & 3.86 & 11.58 & 75.5 \\
 Eq.~(\ref{Eq20}) & 100  & 3.94 & 11.82  & 45 \\
\hline
\end{tabular}
\end{table}

Note that different methods require a different number of the scattering matrix calculations per iteration. For instance, Newton's method and the method of Eq.~(\ref{Eq15}) calculate the first derivative of the scattering matrix, requiring two scattering matrix calculations per iteration, whereas Halley's method and the methods of Eqs.~(\ref{Eq19}),~(\ref{Eq20}) proposed here calculate the second derivative, requiring three scattering matrix calculations per iteration. Therefore, for the comparison purposes, what should be taken into consideration in the first place is the magnitude of $n_{\rm s-calc}$ that defines the total number of the scattering matrix calculations. At the same time, in each iteration, the scattering matrices can be calculated independently, or in parallel. Thus, for parallel implementation of the iterative methods, it is the magnitude of $n_{\rm iter}$ which turns out to be of greater importance.

Table~\ref{tab:table2} was calculated for the truncation order $N = 21$. It can be seen from Table~\ref{tab:table2} that among the ``scalar'' iterative methods for solving Eqs.~(\ref{Eq10}) and~(\ref{Eq11}), Halley's method requires the smallest number of iterations. This is due to the fact that Halley's method has cubic convergence, whereas Newton's method converges with order two~\cite{r31}. Note also that Newton's method shows better convergence for Eq.~(\ref{Eq10}) than for Eq.~(\ref{Eq11}).

In terms of the number of the scattering matrix calculations, the matrix methods of Eqs.~(\ref{Eq15}),~(\ref{Eq19}),~(\ref{Eq20}) possess a higher rate of convergence when compared with the scalar methods. It is seen from Table~\ref{tab:table2} that the methods~(\ref{Eq19}) and~(\ref{Eq20}) proposed here are able to find the poles within a smaller number of iterations among all the methods discussed in this paper. In particular, the average number of iterations for the method~(\ref{Eq19}) is nearly twice as small as that for the method~(\ref{Eq15}). At the same time, the average number of iterations required to calculate the scattering matrix by the method~(\ref{Eq19}) is just 1.31 times smaller than for the method of Eq.~(\ref{Eq15}). The comparison of the methods~(\ref{Eq19}) and~(\ref{Eq20}) shows that the latter approach based on the assumption of a single resonance in the vicinity of the initial pole guess requires on the average a 1.22 times larger number of iterations than a more general method~(\ref{Eq19}).

Let us analyze how the methods under discussion operate at a larger truncation order. Table~\ref{tab:table3} suggests that at $N = 61$ the convergence of the method~(\ref{Eq15}) and Newton's method when solving Eq.~(\ref{Eq10}) is deteriorated: with the pole finding rate being 78\% for the former method and 83\% for the latter. When $N = 201$ (Table~\ref{tab:table4}), both the method~(\ref{Eq15}) and Newton's method for solving Eq.~(\ref{Eq10}) become inoperative. When used to solve Eq.~(\ref{Eq11}), Newton's and Halley's methods are converging, although requiring by about 30\% more iterations  when compared with the methods~(\ref{Eq19}) and~(\ref{Eq20}). The same is true regarding for the number of the scattering matrix calculations. It is noteworthy that the methods~(\ref{Eq19}) and~(\ref{Eq20}) converge in one hundred percent of the cases.

In addition to the rate of convergence, an important characteristic of iterative methods is the shape of the basins of attraction. With regard to the calculation of the poles, basin of attraction of the pole $\omega_{\rm p}$ is the set of points from which the iterative method converges to $\omega_{\rm p}$. Ideally, one would like the iterative method to converge to the nearest pole. In this case, if we vary some parameter of the grating, we will be able to calculate the corresponding variation of the mode eigenfrequency. Hence, we can easily calculate the functional dependence of the eigenfrequency on some parameter (for example we can calculate the dispersion curves by varying the wavenumber $k_{x,0}$). Unfortunately, the Newton's method (as well as the Halley's method) is characterized by the fractal-shaped basins of attraction~\cite{r33}.

\begin{figure*}[t]
\centering
\includegraphics{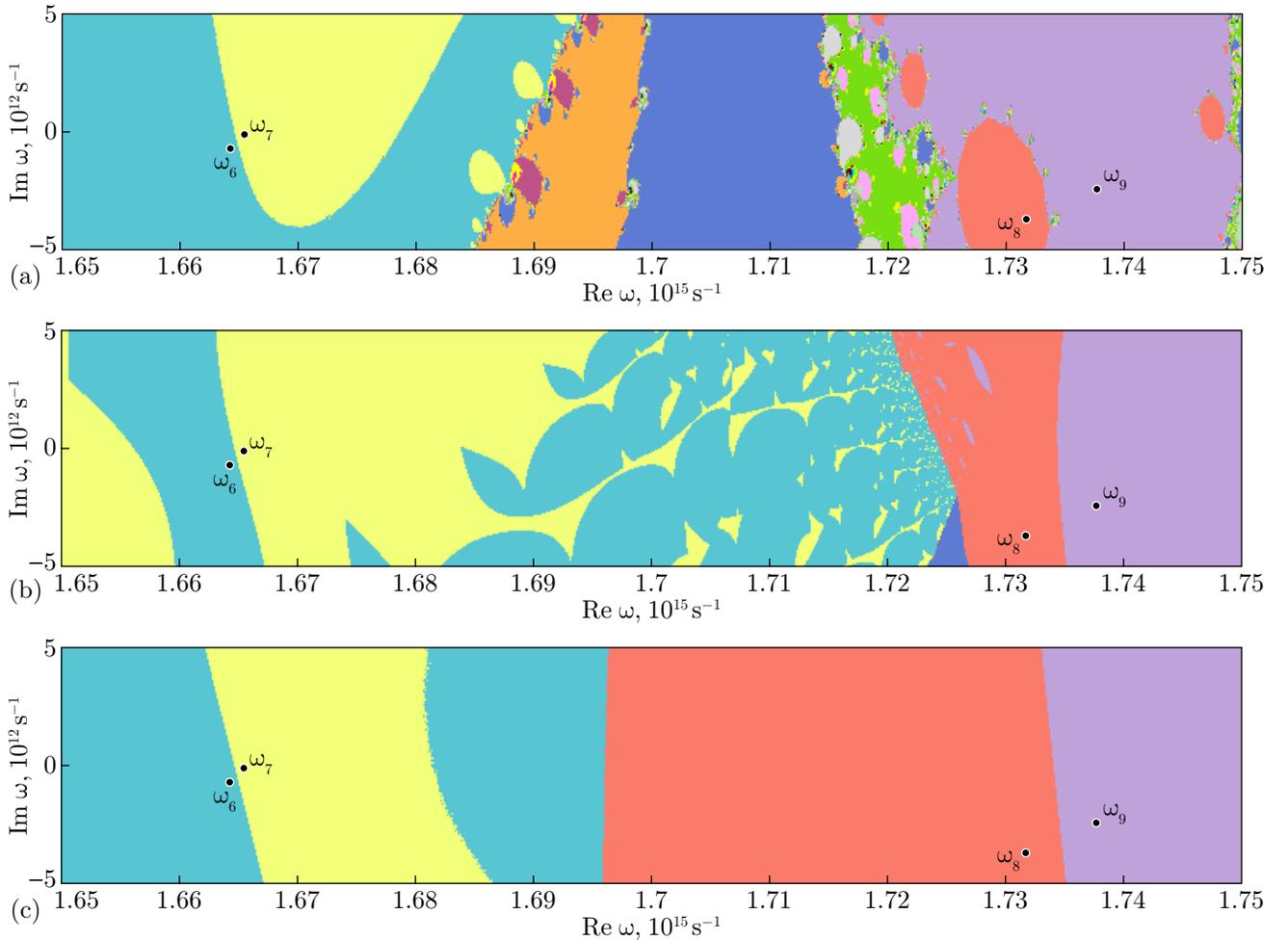}
\caption{\label{fig:fig2}(Color online) Basins of attraction (different colors denote different attraction poles): (a)~Newton for Eq.~\eqref{Eq10}; (b)~Eq.~\eqref{Eq15}; (c)~Eq.~\eqref{Eq19}.}
\end{figure*}

Fig.~\ref{fig:fig2} shows the basins of attraction for $\operatorname{Re}\omega_0 \in [1.65\times 10^{15} \; {\rm s}^{-1} ;1.75\times 10^{15} \; {\rm s}^{-1} ]$, $\operatorname{Im}\omega_0 \in [-5\times 10^{12} \; {\rm s}^{-1} ;5\times 10^{12} \; {\rm s}^{-1} ]$ region. We used the value of $N=21$ as the truncation order. According to Fig.~\ref{fig:fig2} the basins of attraction for the proposed method~(\ref{Eq19}) have rather regular shape while other methods give pronounced fractal shape.

As the numerical characteristic of the shape of the basins of attraction consider the percentage of the starting points from which the method converges to the nearest pole. This value ($n_{\rm near}$) presented in the last column of Tables~\ref{tab:table2}--\ref{tab:table4} has to be close as possible to 100\%. As follows from Tables~\ref{tab:table2}--\ref{tab:table4}, the proposed method~(\ref{Eq19}) in the most cases (70--90\%) converges to the nearest pole. At the same time for the other methods the value of $n_{\rm near}$ ranges from 20 to 60\%.

Tables~\ref{tab:table2}--\ref{tab:table4} demonstrate that the performance of the considered methods decreases with increasing the dimension of the scattering matrix. It must be noted that in all above-mentioned methods except method~(\ref{Eq15}) it is possible to use only a part of the large-dimension scattering matrix. For example, we can use the central part of size $84\times84$ from the large-dimension scattering matrix computed at the truncation order $N=201$. In this case, it is expected that the performances of the matrix methods of Eqs.~(\ref{Eq19}),~(\ref{Eq20}) as well as Newton's (Halley's) method when solving Eqs.~(\ref{Eq10}),~(\ref{Eq11}) will be close to the values given in Table~\ref{tab:table2}.

\section{Conclusion}
Summing up, a new iterative technique for calculating the scattering matrix poles has been proposed in this work. Taking account of the scattering matrix form in the pole vicinity, the method relies upon solving the matrix equations through the use of a matrix decompositions. Using the same mathematical approach, a Cauchy-integral-based method that enables all the poles in a designed region to be calculated has been described. The calculation results for the modes supported by metal-dielectric diffraction grating have shown that the iterative method proposed has the high rate of convergence and is numerically stable for large-dimension scattering matrices. The proposed method converges to 30--60 \% faster compared to known methods. An important advantage of the proposed method is that it converges (in the most cases) to the nearest pole on the complex plane.

Note that the methods discussed in the present article treat the scattering matrix as a function of complex frequency~$\omega$. However, all the methods remain valid if the scattering matrix is treated as a function of the incident light wave number $k_{x,0}$~\cite{r25} or as a function of the $z$-component of the wave-vector of a diffraction order ($k_{z,n}$).
In the case of complex $\omega$, the method can be used to explain the resonances in the transmittance (reflectance) spectrum of the structure.
The use of complex $k_{x,0}$ makes the proposed methods an effective tool to calculate the guided or leaky modes of the grating waveguides.
The complex $k_{z,n}$ approach is employed to describe the modes in the vicinity of Rayleigh frequencies of the diffraction grating~{\cite{r26,r27}}. 

Moreover, the considered methods are not limited to the optical scattering but also can be applied to the problems of acoustical and quantum mechanical scattering.


%

\appendices

\section{\label{appendixA}Solving a system of matrix equations $A=LXR, B=LYR$}
Given matrices $A,B\in \mathbb{C}^{n\times n}$ that take the form: 
\begin{equation}
\label{EqA1}
\left\{
\begin{aligned}
A&=LXR; \\ 
B&=LYR, 
\end{aligned}
\right.
\end{equation} 
where $L\in \mathbb{C}^{n\times r}$, $R\in \mathbb{C}^{r\times n}$ are some unknown matrices, $X\in \mathbb{C}^{r\times r}$, and $Y\in \mathbb{C}^{r\times r}$ are unknown diagonal matrices, with the value of $r$ being also unknown. It is required to find the relation between the matrices $X$ and $Y$. Note that systems of this type are encountered when finding the poles of Linear Time Invariant (LTI) systems~\cite{r14,r15}.

Consider a method for solving the system~(\ref{EqA1}) which is based on the singular value decomposition~\cite{r34}. First, we find the rank $r$ of the matrix $A$. To do so, we calculate its singular value decomposition: $A=U\Sigma V^{\dagger}$ (here $V^{\dagger}$ denotes the conjugate transpose of $V$). The rank $r$ is defined by the amount of nonzero singular numbers. In practice, it may occur that the $A$ matrix is noised. In this case, $r$ is chosen to be equal to the amount of singular numbers larger than a threshold value.
Knowing the rank of $A$, the compact singular value decomposition is given by 
\begin{equation}
\label{EqA2}
A = U_r \Sigma_r V_r^{\dagger}, 
\end{equation}
where $\Sigma_r \in \mathbb{R}^{r\times r}$ is a diagonal matrix composed of $r$ largest singular numbers of the $A$ matrix; $U_r ,V_r \in \mathbb{C}^{n\times r}$ are the matrices of the corresponding left and right singular vectors. If the matrix is noised, the equality in Eq.~(\ref{EqA2}) becomes approximate. In this case Eq.~(\ref{EqA2}) represents the rank-$r$ approximation of matrix $A$.

Note that both $U$ and $V$ are unitary, hence $U_r^{\dagger} U_r = V_r^{\dagger} V_r = I$.
Multiplying the equations in~(\ref{EqA1}) by $U_r^{\dagger}$ on the left and by $V_r$ on the right, we obtain:
\begin{equation}
\label{EqA3}
\left\{
\begin{array}{l}
(U_r^{\dagger} L)X(R V_r ) = (U_r^{\dagger} U_r )\Sigma _r (V_r^{\dagger} V_r )=\Sigma _r ; \\
(U_r^{\dagger} L)Y(R V_r ) = U_r^{\dagger} B V_r.
\end{array}
\right.
\end{equation} 

Note that $\operatorname{rank}\Sigma_r =\operatorname{rank}A=\operatorname{rank}U_r = \operatorname{rank}V_r = r$, hence both $U_r^{\dagger} L$ and $R V_r$ are square and invertible matrices from $\mathbb{C}^{r\times r}$. Multiplying the second equation in~(\ref{EqA3}) by $\Sigma_r^{-1}$ on the right, we get:
\begin{equation}
\label{EqA4}
(U_r^{\dagger} L)YX^{-1} (U_r^{\dagger} L)^{-1} = U_r^{\dagger} B V_r \Sigma _r^{-1}.
\end{equation} 
Equation~(\ref{EqA4}) can be treated as the eigendecomposition of the matrix on the right-hand side of the equation. Thus, the $YX^{-1}$ matrix can be explicitly given by
\begin{equation}
\label{EqA6}
YX^{-1} = \operatorname{diag}\operatorname{eig}(U_r^{\dagger} BV_r \Sigma _r^{-1} ).
\end{equation}

Consider an important particular case when the matrices $A$ and $B$ are of unit rank, i.e. $r=1$. In this case, the vector $U_1$ is the eigenvector of the matrix $A$ and the vector $V_1$ is the eigenvector of the matrix $A^{\dagger}$. Hence, the $A$ matrix can be represented as
\begin{equation}
\label{EqA7}
A=U_{1} \frac{\max\operatorname{eig}\, A}{V_1^{\dagger} U_1 } V_1^{\dagger},
\end{equation} 
where $\max \operatorname{eig}A$ is the only non-zero eigenvalue of the matrix $A$. The vectors $U_1 $, $V_1$ are also the eigenvectors of the matrices $B$ and $B^{\dagger}$, respectively. Thus,
\begin{equation}
\label{EqA8}
B=U_{1} \frac{\max\operatorname{eig}\, B}{V_1^{\dagger} U_1 } V_1^{\dagger}.
\end{equation} 

Substituting~(\ref{EqA7}) and~(\ref{EqA8}) into~(\ref{EqA6}) yields:
\begin{equation}
\label{EqA9}
y/x=\frac{\max\operatorname{eig}\, B}{\max\operatorname{eig}\, A}.
\end{equation} 

Note that because in practice the matrices $A$ and $B$ can be noised, the maximal-modulus eigenvalues should be used in Eq.~(\ref{EqA9}).

\section{\label{appendixB}An example of closely located poles of the matrices $S(\omega)$ and $S^{-1}(\omega)$}
In this appendix, we discuss a specific diffraction grating for which the iterative method in Eq.~(\ref{Eq15}) fails to converge already at a small value of the truncation order. Let us consider a silver diffraction grating (period $d=1000 \; {\rm nm}$, height $h_{\rm gr} = 10 \; {\rm nm}$, slit width $a=990\; {\rm nm}$) on a silver substrate.

Analyze how the method of Eq.~(\ref{Eq15}) performs for the truncation order $N=21$. We are interested in the pole with the frequency $\omega_{\rm p} \approx 1.859\times 10^{15} -1.613\times 10^{12} \; {\rm s}^{-1}$. The peculiarity of this structure is that the scattering matrix determinant turns to zero in the vicinity of $\omega_{\rm p} $, at $\omega \approx 1.863\times 10^{15} -1.465\times 10^{12} \; {\rm s}^{-1} $.

Let the initial guess be $\omega_0 =1.86\times 10^{15} \; {\rm s}^{-1} $. With this initial guess, the proposed iterative method~(\ref{Eq20}) converges within two iterations; whereas the iterative method~(\ref{Eq15}) converges within 5 iterations. However, for a bit worse initial guess of $\omega_0=1.85\times 10^{15} \; {\rm s}^{-1} $, while the proposed iterative method~(\ref{Eq20}) is still converging, the method~(\ref{Eq15}) fails to converge. The reason why the method~(\ref{Eq15}) fails to converge is that $S^{-1}(\omega)$ has a pole and, consequently, the convergence radius of its Taylor series in Eq.~(\ref{Eq13}) is small.


\ifCLASSOPTIONcaptionsoff
  \newpage
\fi



\bibliographystyle{IEEEtran}
\bibliography{bykov}
%



%

%
 \vfill 
\begin{IEEEbiographynophoto}{Dmitry A. Bykov}
graduated with honors (2009) from the Samara State Aerospace University (SSAU), majoring in Applied Mathematics and Computer Science. 
He received his Candidate of Physics and Mathematics (2011) degree from Samara State Aerospace University. 
Currently he is a researcher in the Diffractive Optics Laboratory of the Image Processing Systems Institute of the RAS. 
His current research interests include nanophotonics, magneto-optics of nanostructured materials, plasmonics and grating theory.
\end{IEEEbiographynophoto}

\begin{IEEEbiographynophoto}{Leonid L. Doskolovich}
graduated with honors (1989) from the S.P. Korolyov Kuibyshev Aviation Institute (presently, Samara State Aerospace University (SSAU)), 
majoring in Applied Mathematics. 
He received his Doctor of Physics and Mathematics (2001) degree from Samara State Aerospace University. 
Currently he is a leading researcher of the Image Processing Systems Institute of the RAS, professor at SSAU's Technical Cybernetics department. Current research interests include diffractive optics, laser information technologies, nanophotonics.
\end{IEEEbiographynophoto}





\end{document}